\documentclass{llncs}
\usepackage{cs_techrpt_cover}
\usepackage{graphicx}
\usepackage{amssymb}
\usepackage{eucal}
\usepackage[pdfpagemode=None]{hyperref}

\newcounter{verbline}
\makeatletter
\def\verblineinput#1{
\begingroup
\setcounter{verbline}{0} \fontsize{10pt}{11pt}
\def\par{\leavevmode\null\@@par}
\obeylines \frenchspacing\@vobeyspaces \@makeother\$\@makeother\&\@makeother\#
\@makeother\^\@makeother\_\@makeother\~ \@makeother\{\@makeother\}\@makeother\\
\@makeother\%\tt\@noligs\input{#1}\endgroup} \makeatother

\def\R{\mathop{\rm I\kern-1.5pt R}}
\def\N{\mathop{\rm I\kern-1.5pt N}}

\def\bbbkl{{[\![}}
\def\bbbkr{{]\!]}}
\def\cc{{\cal U}}
\def\crel{{\cal R}}

\def\cx{{\cal X}}

\title{CrocoPat 2.1 Introduction and Reference Manual\thanks{%
  This research was supported in part by
  the NSF grants CCR-0234690 and ITR-0326577,
  and the DFG grant BE~1761/3-1.
       }}

\institute{
   Electronics Research Laboratory \\
   Department of Electrical Engineering and Computer Sciences \\
   College of Engineering \\
   University of California at Berkeley \\
   Berkeley, CA 94720-1770, U.S.A. \\
   \email{beyer@eecs.berkeley.edu}\\[5mm]
\and
   Software Systems Engineering Research Group\\
   Department of Computer Science\\
   Brandenburg University of Technology at Cottbus\\
   PO Box 10 13 44, 03013 Cottbus, Germany\\
   \email{an@informatik.tu-cottbus.de}
}

\date{July 2004}
\reportnumber{04-1338}
\useuglyformat

\begin{document}

\author{Dirk Beyer \and Andreas Noack}
\makecover

\phantom{x}

\author{Dirk Beyer\inst{1} \and Andreas Noack\inst{2}}
\maketitle

\pagestyle{plain}

\begin{abstract}
CrocoPat is an efficient, powerful and easy-to-use tool for manipulating
relations of arbitrary arity, including directed graphs.  This manual provides
an introduction to and a reference for CrocoPat and its programming
language~RML. It includes several application examples, in particular from the
analysis of structural models of software systems.
\end{abstract}

\sloppy


\section{Introduction} \label{s:intro}

CrocoPat is a tool for manipulating relations, including directed
graphs (binary relations).  CrocoPat is powerful, because it
manipulates relations of arbitrary arity; it is efficient in terms
of time and memory, because it uses the data structure binary
decision diagram (BDD,~\cite{Bryant:1986,Bryant:1992}) for the
internal representation of relations; it is fairly easy to use,
because its language is simple and based on the well-known
predicate calculus; and it is easy to integrate with other tools,
because it uses the simple and popular Rigi Standard Format (RSF)
as input and output format for relations. CrocoPat is free
software (released under LGPL) and can be obtained from
\href{http://www.software-systemtechnik.de/CrocoPat}{\tt
http://www.software-systemtechnik.de/CrocoPat}.

{\bf Overview.} CrocoPat is a command line tool which interprets programs
written in the Relation Manipulation Language (RML).  Its inputs are an RML
program and relations in the Rigi Standard Format (RSF), and its outputs are
relations in RSF and other text produced by the RML program.  The programming
language RML, and the input and output of relations from and to RSF files are
introduced with the help of many examples in Section~\ref{s:lanintro}.
Section~\ref{s:advanced} describes advanced programming techniques, in
particular for improving the performance of RML programs and for circumventing
limitations of RML. The manual concludes with references of CrocoPat's command
line options in Section~\ref{s:pat}, of RSF in Section~\ref{s:rsf}, and of RML
in Section~\ref{s:lan}.  The RML reference includes a concise informal
description of the semantics, and a formal description of the syntax and the
core semantics.

{\bf Applications.} CrocoPat was originally developed for analyzing graph
models of software systems, and in particular for finding patterns in such
graphs~\cite{BeyerEtAl:2003}. Existing tools were not appropriate for this
task, because they were limited to binary relations (e.g.
Grok~\cite{Holt:1998}, RPA \cite{FeijsEtAl:1998}, and
RelView~\cite{BerghammerEtAl:2002}), or consumed too much time or memory (e.g.
relational database management systems and Prolog interpreters). Applications
of graph pattern detection include
\begin{itemize}
\item the detection of implementation
patterns~\cite{RichWills:1990,HarandiNing:1990,Hartman:1991,Quilici:1994},
object-oriented design patterns (Section~\ref{s:pattern},
\cite{MendelzonSametinger:1995,KraemerPrechelt:1996,AntoniolEtAl:1998,KellerEtAl:1999,NiereEtAl:2002}),
and architectural styles~\cite{Holt:1996},
\item the detection of potential design problems (Section~\ref{s:pattern},
\cite{MendelzonSametinger:1995,SefikaEtAl:1996a,FeijsEtAl:1998,KazmanBurth:1998,Ciupke:1999,FahmyHolt:2000}),
\item the inductive inference of design
patterns~\cite{ShullEtAl:1996,TonellaAntoniol:1999},
\item the identification of code clones~\cite{Krinke:2001},
\item the extraction of scenarios from models of source
code~\cite{WuEtAl:2002}, and
\item the detection of design problems in databases~\cite{Blaha:2004}.
\end{itemize}

The computation of transitive closures of graphs -- another particular strength
of CrocoPat -- is not only needed for the detection of some of the above
patterns, but has also been applied for dead code detection and change impact
analysis~\cite{ChenEtAl:1998,FeijsEtAl:1998}.  Computing and analyzing the
difference between two graphs supports checking the conformance of the as-built
architecture to the as-designed architecture
\cite{SefikaEtAl:1996a,FeijsEtAl:1998,MensWuyts:1999,FahmyHolt:2000,MurphyEtAl:2001},
and studying the evolution of software systems between different versions.
Calculators for relations have also been used to compute views of systems on
different levels of abstraction by lifting and lowering
relations~\cite{FeijsEtAl:1998,FahmyHolt:2000}, and to calculate software
metrics (Section~\ref{s:numerical},
\cite{MendelzonSametinger:1995,KullbachWinter:1999}).

Although we are most familiar with potential applications in the analysis of
software designs, we are confident that CrocoPat can be beneficial in many
other areas. For example, calculators for relations have been used for program
analyses like points-to analysis~\cite{BerndlEtAl:2003}, and for the
implementation of graph algorithms~(Section~\ref{s:control},
\cite{BehnkeEtAl:1997}).

\section{RML Tutorial} \label{s:lanintro}

This section introduces RML, the programming language of CrocoPat, on examples.
The core of RML are relational expressions based on first-order predicate
calculus, a well-known, reasonably simple, precise and powerful language.
Relational expressions are explained in Subsection~\ref{s:re}, and additional
examples involving relations of arity greater than two are given in
Subsection~\ref{s:pattern}.  Besides relational expressions, the language
includes control structures, described in Subsection~\ref{s:control}, and
numerical expressions, described in Subsection~\ref{s:numerical}.  The input
and output of relations is described in Subsection~\ref{s:output}.  A more
concise and a more formal specification of the language can be found in
Section~\ref{s:lan}.

Although the main purpose of this section is the introduction of the language,
some of the application examples may be of interest by themselves. In
Subsection~\ref{s:control}, simple graph algorithms are implemented, and in the
Subsections~\ref{s:pattern} and~\ref{s:numerical} the design of object-oriented
software systems is analyzed.

\subsection{Relational Expressions} \label{s:re}

This subsection introduces relational expressions using relationships between
people as example. Remember that $n$-ary relations are sets of ordered
$n$-tuples.  In this subsection, we will only consider the cases $n = 1$ (unary
relations) and $n = 2$ (binary relations, directed graphs).  CrocoPat
manipulates tuples of strings, thus unary relations in CrocoPat are sets of
strings, and binary relations in CrocoPat are sets of ordered pairs of strings.

{\bf Adding Elements.} The statement
\begin{verbatim}
Male("John");
\end{verbatim}
expresses that {\tt John} is male. (In some languages, e.g. the logic
programming language Prolog~\cite{ClocksinMellish:2003}, such statements are
called facts.) It adds the string {\tt John} to the unary relation {\tt Male}.
Because each relation variable initially contains the empty relation, {\tt
John} is so far the only element of the set {\tt Male}. An explicit declaration
of variables is not necessary. However, variables should be defined (i.e.,
assigned a value) before they are first used, otherwise CrocoPat prints a
warning.
\begin{verbatim}
Male("Joe");
\end{verbatim}
adds the string {\tt Joe} to the set {\tt Male}, such that it now has two
elements. Similarly, we can initialize the variable {\tt Female}:
\begin{verbatim}
Female("Alice");
Female("Jane");
Female("Mary");
\end{verbatim}

To express that the {\tt John} and {\tt Mary} are the
parents of {\tt Alice} and {\tt Joe}, and {\tt Joe} is the father of {\tt
Jane}, we create a binary relation variable {\tt ParentOf} which contains the
five parent-child pairs:
\begin{verbatim}
ParentOf("John", "Alice");
ParentOf("John", "Joe");
ParentOf("Mary", "Alice");
ParentOf("Mary", "Joe");
ParentOf("Joe", "Jane");
\end{verbatim}

{\bf Assignments.} The following statement uses an {\em attribute}~{\tt x} to
assign the set of {\tt Joe}'s parents to the set {\tt JoesParent}:
\begin{verbatim}
JoesParent(x) := ParentOf(x, "Joe");
\end{verbatim}
Now {\tt JoesParent} contains the two elements {\tt John} and {\tt Mary}. As
another example, the following assignment says that {\tt x} is a child of {\tt
y} if and only if {\tt y} is a parent of {\tt x}:
\begin{verbatim}
ChildOf(x,y) := ParentOf(y,x);
\end{verbatim}
{\tt John} is the father of a person if and only if he is the parent of this
person.  The same is true for {\tt Joe}:
\begin{verbatim}
FatherOf("John", x) := ParentOf("John", x);
FatherOf("Joe", x) := ParentOf("Joe", x);
\end{verbatim}
Because the scope of each attribute is limited to one statement, the attribute
in the first statement and the attribute in the second statement are different,
despite of their equal name~{\tt x}.

{\bf Basic Relational Operators.} The relation {\tt FatherOf} can be described
more concisely: {\tt x} is father of~{\tt y} if and only if {\tt x} is a parent
of {\tt y} and {\tt x} is male:
\begin{verbatim}
FatherOf(x,y) := ParentOf(x,y) & Male(x);
\end{verbatim}
Of course, we can define a similar relation for female parents:
\begin{verbatim}
MotherOf(x,y) := ParentOf(x,y) & Female(x);
\end{verbatim}

Besides the operator {\em and} ({\tt \&}), another important operator is {\em
or} ({\tt |}). For example, we can define the {\tt ParentOf} relation in terms
of the relations {\tt MotherOf} and {\tt FatherOf}: {\tt x}~is a parent of~{\tt
y} if and only if {\tt x}~is the mother or the father of~{\tt y}:
\begin{verbatim}
ParentOf(x,y) := MotherOf(x,y) | FatherOf(x,y);
\end{verbatim}

{\bf Quantification.} Parents are people who are a parent of another person.
More precisely, {\tt x} is a parent if and only if there {\em exists} ({\tt
EX}) a {\tt y} such that {\tt x} is a parent of {\tt y}.
\begin{verbatim}
Parent(x) := EX(y, ParentOf(x,y));
\end{verbatim}
Now the set {\tt Parent} consists of {\tt John}, {\tt Mary}, and {\tt Joe}.
There is also an abbreviated notation for existential quantification which is
similar to anonymous variables in Prolog and functional programming languages:
\begin{verbatim}
Parent(x) := ParentOf(x,_);
\end{verbatim}
With the operator {\em not} ({\tt !}), we can compute who has no children:
\begin{verbatim}
Childless(x) := !EX(y, ParentOf(x,y));
\end{verbatim}
Equivalently, {\tt x} childless if {\em for all} ({\tt FA}) {\tt y} holds that
{\tt x} is not a parent of {\tt y}:
\begin{verbatim}
Childless(x) := FA(y, !ParentOf(x,y));
\end{verbatim}
In both cases, the set {\tt Childless} contains {\tt Alice} and {\tt Jane}.

{\bf Transitive Closure.} To compute the grandparents of a person we have to
determine the parents of his or her parents:
\begin{verbatim}
GrandparentOf(x,z) := EX(y, ParentOf(x,y) & ParentOf(y,z));
\end{verbatim}
Now {\tt GrandparentOf} contains the two pairs ({\tt John}, {\tt Jane}) and
({\tt Mary}, {\tt Jane}).  To find out all ancestors of a person, i.e. parents,
parents of parents, parents of parents of parents, etc., we have to apply the
above operation (which is also called composition) repeatedly until the fixed
point is reached, and unite the results.  The transitive closure operator {\tt
TC} does exactly this:
\begin{verbatim}
AncestorOf(x,z) := TC(ParentOf(x,z));
\end{verbatim}
The resulting relation {\tt AncestorOf} contains any pair from {\tt ParentOf}
and {\tt GrandparentOf}. (It also contains grand-grandparents etc., but there
are none in this example.) The transitive closure operator {\tt TC} can only be
applied to binary relations.

{\bf Predefined Relations, the Universe.} The relations {\tt FALSE} and {\tt
TRUE} are predefined.  {\tt FALSE} is the empty relation, and {\tt TRUE} is the
full relation. More precisely, there is one predefined relation {\tt FALSE} and
one predefined relation {\tt TRUE} for every arity.  In particular, there is
also a 0-ary relation {\tt FALSE()}, which is the empty set, and a 0-ary
relation {\tt TRUE()}, which contains only~$()$ (the tuple of length~$0$).
Intuitively, these 0-ary relations can be used like Boolean literals.  By the
way, the statement
\begin{verbatim}
Male("John");
\end{verbatim}
is an abbreviation of the assignment
\begin{verbatim}
Male("John") := TRUE();
\end{verbatim}

The result of {\tt TRUE(x)} is the so-called {\em universe}.  The universe
contains all string literals that appear in the input RSF stream (if there is
one, see Subsection~\ref{s:output}) and on the left hand side of assignments in
the present RML program.  For example, the string literals used on the left
hand side of the assignments in the examples of this subsection are {\tt
Alice}, {\tt Jane}, {\tt Joe}, {\tt John}, and {\tt Mary}, so the set {\tt
TRUE(x)} contains these five elements. See Subsection~\ref{s:universe} for more
information on the universe.

The binary relations {\tt =}, {\tt !=}, {\tt <}, {\tt <=}, {\tt >}, and {\tt
>=} for the lexicographical order of the strings in the universe are also predefined.
For example, siblings are two {\em different} people who have a common parent:
\begin{verbatim}
SiblingOf(x,y) := EX(z, ParentOf(z,x) & ParentOf(z,y)) & !=(x,y);
\end{verbatim}
The infix notation is also available for binary relations, so the expression
{\tt !=(x,y)} can also be written as~{\tt x!=y}.  Note that the predefined
relations, like any other relation, are restricted to the universe. Thus the
expression {\tt "A"="A"} yields {\tt FALSE()} if (and only if) the string {\tt
A} is not in the universe.

Further relational expressions are provided to match POSIX extended regular
expressions~\cite[Section~9.4]{IEEE:2001}.  These relational expressions start
with the character {\tt @}, followed by the string for the regular expression.
For example,
\begin{verbatim}
StartsWithJ(x) := @"^J"(x);
\end{verbatim}
assigns to the set {\tt StartsWithJ} the set of all strings in the universe
that start with the letter {\tt J}, namely {\tt Jane}, {\tt Joe}, and {\tt
John}. A short overview of the syntax of regular expressions is given in
Subsection~\ref{s:syntax}.

{\bf Boolean Operators.} Two relations can be compared with the operators {\tt
=}, {\tt !=}, {\tt <}, {\tt <=}, {\tt >}, or {\tt >=}.  Because such
comparisons evaluate to either {\tt TRUE()} or {\tt FALSE()}, they are called
Boolean expressions. For example,
\begin{verbatim}
GrandparentOf(x,y) < AncestorOf(x,y)
\end{verbatim}
yields {\tt TRUE()}, because {\tt GrandparentOf} is a proper subset of {\tt
AncestorOf}.  However,
\begin{verbatim}
GrandparentOf(x,y) = AncestorOf(x,y)
\end{verbatim}
yields {\tt FALSE()}, because the two relations are not equal.

The six comparison operators should not be confused with the six predefined
relations for the lexicographical order. The operators take two relations as
parameters, while the predefined relations take strings or attributes as
parameters.

\subsection{Input and Output of Relations} \label{s:output}

{\bf File Format RSF.} CrocoPat reads and writes relations in Rigi Standard
Format (RSF,~\cite[Section~4.7.1]{Wong:1998}). Files in RSF are human-readable,
can be loaded into and saved from many reverse engineering tools, and are
easily processed by scripts in common scripting languages.

In an RSF file, a tuple of an $n$-ary relation is represented as a line of the
form
\begin{verbatim}
RelationName element1 element2 ... elementn
\end{verbatim}
The elements may be enclosed by double quotes. Because white space serves as
delimiter of the elements, elements that contain white space {\em must} be
enclosed by double quotes. A relation is represented as a sequence of such
lines.  The order of the lines is arbitrary.  An RSF file may contain several
relations.

As an example, the relation {\tt ParentOf} from the previous subsection can be
represented in RSF format as follows:
\begin{verbatim}
ParentOf John Alice
ParentOf John Joe
ParentOf Mary Alice
ParentOf Mary Joe
ParentOf Joe  Jane
\end{verbatim}

{\bf Input.} RML has no input statements.  When CrocoPat is started, it first
reads input relations in RSF from the standard input before it parses and
executes the RML program.  RSF reading can be skipped with the {\tt -e} command
line option. If the input relations are available as files, they can be feeded
into CrocoPat's standard input using the shell operator {\tt <}, as the
following examples shows for the file {\tt ParentOf.rsf}:
\begin{verbatim}
crocopat Prog.rml < ParentOf.rsf
\end{verbatim}
The end of the input data is recognized either from the end of file character
or from a line that starts with the dot ({\tt .}) character.  The latter is
sometimes useful if RSF input is feeded interactively.

If the above RSF data is used as input, then at the start of the program the
binary relation variable {\tt ParentOf} contains the five pairs, and the
universe contains the five string literals {\tt Alice}, {\tt Jane}, {\tt Joe},
{\tt John}, and {\tt Mary} (and additionally all string literals that appear on
the left hand side of assignments in the program.)

{\bf Output.} The {\tt PRINT} statement outputs relations in RSF format to the
standard output. For example, running the program
\begin{verbatim}
ParentOf("Joe",x) := FALSE(x);
ParentOf(x,"Joe") := FALSE(x);
PRINT ParentOf(x,y);
\end{verbatim}
with the above input data prints to the standard output
\begin{verbatim}
John Alice
Mary Alice
\end{verbatim} The statement
\begin{verbatim}
PRINT ["ParentOf"] ParentOf(x,y);
\end{verbatim}
writes the string {\tt ParentOf} before each tuple, and thus outputs
\begin{verbatim}
ParentOf John Alice
ParentOf Mary Alice
\end{verbatim}
The output can also be appended to a file {\tt ParentOf2.rsf} (which is created
if it does not exist) with
\begin{verbatim}
PRINT ["ParentOf"] ParentOf(x,y) TO "ParentOf2.rsf";
\end{verbatim}
or to stderr with
\begin{verbatim}
PRINT ["ParentOf"] ParentOf(x,y) TO STDERR;
\end{verbatim}
\pagebreak

{\bf Command Line Arguments.} It is sometimes convenient to specify the names
of output files at the command line and not in the RML program.  If there is
only one output file, the standard output can be simply redirected to a file
using the shell operator {\tt >}:
\begin{verbatim}
crocopat Prog.rml < ParentOf.rsf > ParentOf2.rsf
\end{verbatim}

An alternative solution (which also works with more than one file) is to pass
command line arguments to the program.  Command line arguments can be accessed
in RML as {\tt \$1}, {\tt \$2}, etc.  For example, when the program
\begin{verbatim}
ChildOf(x,y) := ParentOf(y,x);
PRINT ["Child"] ChildOf(x,$1) TO $1 + ".rsf";
PRINT ["Child"] ChildOf(x,$2) TO $2 + ".rsf";
\end{verbatim}
is executed with
\begin{verbatim}
crocopat IO.rml Joe Mary < ParentOf.rsf
\end{verbatim}
then the first {\tt PRINT} statement writes to the file {\tt Joe.rsf}, and
the second {\tt PRINT} statement writes to {\tt Mary.rsf}.

Command line arguments are not restricted to specifying file names, but can be
used like string literals.  However, in contrast to string literals, command
line arguments are never added to the universe, and thus cannot be used on the
left hand side of relational assignments.

\subsection{Control Structures} \label{s:control}

This subsection introduces the control structures of RML, using algorithms for
computing the transitive closure of a binary relation~{\tt R} as examples.

{\bfseries {\ttfamily WHILE} Statement.} As a first algorithm, the relation
{\tt R} is composed with itself until the fixed point is reached.
\begin{verbatim}
Result(x,y) := R(x,y);
PrevResult(x,y) := FALSE(x,y);
WHILE (PrevResult(x,y) != Result(x,y)) {
    PrevResult(x,y) := Result(x,y);
    Result(x,z) := Result(x,z) | EX(y, Result(x,y) & Result(y,z));
}
\end{verbatim}
The program illustrates the use of the {\tt WHILE} loop, which has the usual
meaning: The body of the loop is executed repeatedly as long as the condition
after {\tt WHILE} evaluates to~{\tt TRUE()}.

{\bfseries {\ttfamily FOR} Statement.} The second program computes the
transitive closure of the relation~{\tt R} using the Warshall algorithm.  This
algorithm successively adds arcs. In the first iteration, an arc $(u,v)$ is
added if the input graph contains the arcs $(u,$~node$_0)$ and $($node$_0, v)$.
In the second iteration, an arc $(u, v)$ is added if the graph that results
from the first iteration contains the arcs $(u,$~node$_1)$ and $($node$_1, v)$.
And so on, for all nodes of the graph (in arbitrary order.)
\begin{verbatim}
Result(x,y) := R(x,y);
Node(x) := Result(x,_) & Result(_,x);
FOR node IN Node(x) {
  Result(x,y) := Result(x,y) | (Result(x,node) & Result(node,y));
}
\end{verbatim}
The program illustrates the use of the {\tt FOR} loop.  The relation after {\tt
IN} must be a unary relation.  The iterator after {\tt FOR} is a string
variable and takes as values the elements of the unary relation in
lexicographical order. Thus, the number of iterations equals the number of
elements of the unary relation.

For the implementation of the transitive closure operator of RML, we
experimented with several algorithms.  An interesting observation in these
experiments was that the empirical complexity of some algorithms for practical
graphs deviated strongly from their theoretical worst case complexity, thus
some algorithms with a relatively bad worst-case complexity were very
competitive in practice.  In our experiments, the first of the above algorithms
was very fast, thus we made it available as operator {\tt TCFAST}.  The
implementation of the {\tt TC} operator of RML is a variant of the Warshall
algorithm.  It is somewhat slower than {\tt TCFAST} (typically about 20 percent
in our experiments), but often needs much less memory because it uses no
ternary relations.

{\bfseries {\ttfamily IF} Statement.} The following example program determines
if the input graph {\tt R} is acyclic, by checking if its transitive closure
contains loops (i.e. arcs from a node to itself):
\begin{verbatim}
SelfArcs(x,y) := TC(R(x,y)) & (x = y);
IF (SelfArcs(_,_)) {
    PRINT "R is not acyclic", ENDL;
} ELSE {
    PRINT "R is acyclic", ENDL;
}
\end{verbatim}

\subsection{Relations of Higher Arity}\label{s:pattern}

In this subsection, relations of arity greater than two are used for finding
potential design patterns and design problems in structural models of
object-oriented programs.  The examples are taken from~\cite{BeyerEtAl:2003}.

The models of object-oriented programs contain the call, containment, and
inheritance relationships between classes. Here containment means that a class
has an attribute whose type is another class. The direction of inheritance
relationships is from the subclass to the superclass. As an example, the source
code
\begin{verbatim}
class ContainedClass {}
class SuperClass {}
class SubClass extends SuperClass {
    ContainedClass c;
}
\end{verbatim}
corresponds to the following RSF file:
\begin{verbatim}
Inherit   SubClass  SuperClass
Contain   SubClass  ContainedClass
\end{verbatim}

\begin{figure}[ht]
   \centering \includegraphics{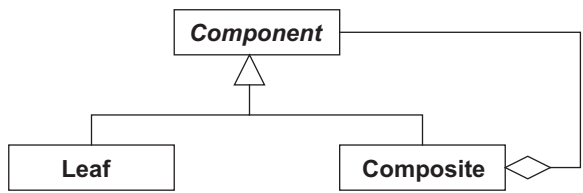}
   {\caption{Composite design pattern
   \label{fig:composite}}}
\end{figure}

{\bf Composite Design Pattern.} Figure~\ref{fig:composite} shows the class
diagram of the Composite design pattern~\cite{GammaEtAl:1995}. To identify
possible instances of this pattern, we compute all triples of a Component
class, a Composite class, and a Leaf class, such that (1) Composite and Leaf
are subclasses of Component, (2) Composite contains an instance of Component,
and (3) Leaf does not contain an instance of Component. The translation of
these conditions to an RML statement is straightforward:
\begin{verbatim}
CompPat(component, composite, leaf) :=   Inherit(composite, component)
                                       & Contain(composite, component)
                                       & Inherit(leaf, component)
                                       & !Contain(leaf, component);
\end{verbatim}

{\bf Degenerate Inheritance.} When a class {\tt C} inherits from another class
{\tt A} directly and indirectly via a class {\tt B}, the direct inheritance is
probably redundant or even misleading.  The following statement detects such
patterns:
\begin{verbatim}
DegInh(a,b,c) :=   Inherit(c,b)
                 & Inherit(c,a)
                 & TC(Inherit(b,a));
\end{verbatim}

{\bf Cycles.} To understand an undocumented class, one has to understand all
classes it uses. If one of the (directly or indirectly) used classes is the
class itself, understanding this class is difficult.  All classes that
participate in cycles can be found using the transitive closure operator, as
shown in Subsection~\ref{s:control}.  However, in many large software systems
hundreds of classes participate in cycles, and it is tedious for a human
analyst to find the actual cycles in the list of these classes. In our
experience, it is often more useful to detect cycles in the order of ascending
length.  As a part of such a program, the following statements detects all
cycles of length~$3$.
\begin{verbatim}
Use(x,y) := Call(x,y) | Contain(x,y) | Inherit(x,y);
Cycle3(x,y,z) := Use(x,y) & Use(y,z) & Use(z,x);
Cycle3(x,y,z) := Cycle3(x,y,z) & (x <= y) & (x <= z);
\end{verbatim}
To see the purpose of the third statement, consider three classes {\tt A}, {\tt
B}, and {\tt C} that form a cycle.  After the second statement, the relation
variable {\tt Cycle3} contains three representatives of this cycle: $(${\tt
A},~{\tt B},~{\tt C}$)$, $(${\tt B},~{\tt C},~{\tt A}$)$ and $(${\tt C},~{\tt
A},~{\tt B}$)$. The third statement removes two of these representatives from
{\tt Cycle3}, and keeps only the tuple with the lexicographically smallest
class at the first position, namely $(${\tt A},~{\tt B},~{\tt C}$)$.

\subsection{Numerical Expressions} \label{s:numerical}

In this subsection, a software metric is calculated as example for the use of
numerical expressions in RML programs.  Therefore, we extend the structural
model of object-oriented programs introduced in the previous subsection with a
binary relation {\tt PackageOf}.  This relation assigns to each package the
classes that it contains. (Packages are high-level entities in object-oriented
software systems that can be considered as sets of classes.)

Robert Martin's metric for the instability of a package is defined as $ce /
(ca+ce)$, where $ca$ is the number of classes outside the package that use
classes inside the package, and $ce$ is the number of classes inside the
package that use classes outside the package~\cite{Martin:1997}.
\begin{verbatim}
Use(x,y) := Call(x,y) | Contain(x,y) | Inherit(x,y);
Package(x) := PackageOf(x,_);
FOR p IN Package(x) {
    CaClass(x) := !PackageOf(p,x) & EX(y, Use(x,y) & PackageOf(p,y));
    ca := #(CaClass(x));
    CeClass(x) := PackageOf(p,x) & EX(y, Use(x,y) & !PackageOf(p,y));
    ce := #(CeClass(x));
    IF (ca + ce > 0) {
        PRINT p, " ", ce / (ca+ce), ENDL;
    }
}
\end{verbatim}

\section{Advanced Programming Techniques} \label{s:advanced}

This section describes advanced programming techniques, in particular for
improving efficiency and circumventing language limitations.  The first
subsection explains how to control the memory usage of CrocoPat.  The second
and third subsection describe how relational expressions are evaluated in
CrocoPat, and how to assess and improve the efficiency of their evaluation. The
fourth subsection explains why the universe is immutable during the execution
of an RML program and how to work around this limitation.

\subsection{Controlling the Memory Usage} \label{ss:memory}

CrocoPat represents relations using the data structure binary decision diagram
(BDD,~\cite{Bryant:1986}). When CrocoPat is started, it reserves a fixed amount
of memory for BDDs, which is not changed during the execution of the RML
program. If the available memory is insufficient, CrocoPat exits with the error
message
\begin{verbatim}
Error: BDD package out of memory.
\end{verbatim}

The BDD memory can be controlled with the command line option~{\tt -m},
followed by an integer number giving the amount of memory in MByte.  The
default value is~50.  The actual amount of memory reserved for BDDs is not
infinitely variable, so the specified value is only a rough upper bound of the
amount of memory used.

It can also be beneficial to reserve less memory, because the time used for
allocating memory increases with the amount of memory.  When the manipulated
relations are small or the algorithms are computationally inexpensive, memory
allocation can dominate the overall runtime.

\subsection{Speeding up the Evaluation of Relational Expressions}

This subsection explains how CrocoPat evaluates relational expressions.  Based
on this information, hints for performance improvement are given. Understanding
the subsection requires basic knowledge about BDDs and the impacts of the
variable order on the size of BDDs. An introduction to BDDs is beyond the scope
of this manual, we refer the reader to~\cite{Bryant:1992}.

The attributes in an RML program are called {\em user attributes} in the
following.  For example, the expression {\tt R(x,y)} contains the user
attributes {\tt x} and~{\tt y}.  For the internal representation of relations,
CrocoPat uses a sequence of {\em internal attributes}, which are distinct from
the user attributes.  We call these internal attributes {\tt i1}, {\tt i2},
{\tt i3}, ... For example, the binary relation {\tt R} is internally
represented as a set of assignments to the internal attributes {\tt i1}
and~{\tt i2}.

When the expression {\tt R(x,y)} is evaluated, the internal attributes {\tt i1}
and {\tt i2} are renamed to the user attributes {\tt x} and {\tt y}. Therefore
all BDD nodes of the representation of {\tt R} have to be traversed. Thus, the
time for evaluating the expression {\tt R(x,y)} is at least linear in the
number of BDD nodes of~{\tt R}'s representation.

The order of the internal attributes in the BDD is always {\tt i1}, {\tt i2},
... The order of the user attributes in the BDD may be different in the
evaluation of each statement, because the scope of user attributes is restricted
to one statement. The order of the user attributes in the BDD in the evaluation
of a statement is the order in which CrocoPat encounters the user attributes in
the execution of the statement.  In the example statement
\begin{verbatim}
R(x,z) := EX(y, R(x,y) & R(y,z));
\end{verbatim}
CrocoPat evaluates first {\tt R(x,y)}, then {\tt R(y,z)}, then the conjunction,
then the existential quantification, and finally the assignment.  Therefore,
the order of the user attributes in the BDD is {\tt x}, {\tt y}, {\tt z}.

{\bf Avoid renaming large relations.} The time for the evaluation of the
expression {\tt R(x,y)} is at least linear in the number of BDD nodes in the
representation of~{\tt R}, because all BDD nodes have to be renamed from
internal attributes to user attributes. Usually this effort for renaming does
not dominate the overall runtime, but in the following we give an example where
it does. \pagebreak

Let {\tt R(x,y)} be a directed graph with $n$~nodes. Let the BDD representation
of {\tt R} have $\Theta(n^2)$ BDD nodes (which is the worst case). The
assignment
\begin{verbatim}
Outneighbor(y) := EX(x, R(x,y) & x="node1");
\end{verbatim}
assigns the outneighbors of the graph node {\tt node1} to the set {\tt
Outneighbor}.  The evaluation of {\tt R(x,y)} costs $\Theta(n^2)$ time in this
example, because of the renaming of all nodes.  The ``real computation'',
namely the conjunction and the existential quantification, can be done in
$O(\log n)$ time.  So the renaming dominates the overall time.

The equivalent statement
\begin{verbatim}
Outneighbor(y) := R("node1",y);
\end{verbatim}
is executed in only $O(n)$ time, because the set {\tt R("node1",y)} has $O(n)$
elements, and its BDD representation has $O(n)$ nodes.

{\bf Avoid swapping attributes.} Renaming the nodes of a BDD costs at least
linear time, but can be much more expensive when attributes have to be swapped.
In the statement
\begin{verbatim}
S(x,y,z) := R(y,z) & R(x,y);
\end{verbatim}
the BDD attribute order on the right hand side of the assignment is {\tt y},
{\tt z}, {\tt x}, while the BDD attribute order on the left hand side is {\tt
x}, {\tt y}, {\tt z}. Because the two orders are different, attributes have to
be swapped to execute the assignment. This can be easily avoided by using the
equivalent statement
\begin{verbatim}
S(x,y,z) := R(x,y) & R(y,z);
\end{verbatim}

Of course, swapping attributes can not always be avoided.  However, developers
of RML programs should know that swapping attributes can be expensive, and
should minimize it when performance is critical.

{\bf Ensure good attribute orders.} A detailed discussion of BDD attribute
orders is beyond the scope of this manual (see
e.g.~\cite[Section~1.3]{Bryant:1992} for details), but the basic rule is that
related attributes should be grouped together.  In the two assignment
statements
\begin{verbatim}
S1(v,w,x,y) := R(v,w) & R(x,y);
S2(v,x,w,y) := R(v,w) & R(x,y);
\end{verbatim}
the attributes {\tt v} and {\tt w} are related, and the attributes {\tt x} and
{\tt y} are related, while {\tt v} and {\tt w} are unrelated to {\tt x}
and~{\tt y}.  In {\tt S1}, related attributes are grouped together, but not
in~{\tt S2}.  For many relations~{\tt R}, the BDD representation of~{\tt S1}
will be drastically smaller than the BDD representation of~{\tt S2}.

{\bf Profile.} Information about the number of BDD nodes and the BDD attribute
order of an expression can be printed with {\tt PRINT RELINFO}. For example,
\begin{verbatim}
PRINT RELINFO(R(y,z) & R(x,y));
\end{verbatim}
may output
\begin{verbatim}
Number of tuples in the relation: 461705
Number of values (universe): 6218
Number of BDD nodes: 246986
Percentage of free nodes in BDD package: 1614430 / 1966102 = 82 %
Attribute order: y z x
\end{verbatim}
The first line gives the cardinality of the relation, the second line the
cardinality of the universe, the third line the size of the BDD that represents
the result of the expression, and the fifth line the attribute order in this
BDD.

\subsection{Estimating the Evaluation Time of Relational Expressions} \label{ss:estimation}

Knowledge of the computational complexity of RML's operators is useful to
optimize the performance of RML programs. This subsection gives theoretical
complexity results, but also discusses the limits of their practical
application.

Table~\ref{t:complexity} shows the asymptotic worst case time complexity for
the evaluation of RML's relational operators.  The times do not include the
renaming of internal attributes discussed in the previous subsection, and
the evaluation of subexpressions.  It is assumed that the caches of the BDD
package are sufficiently large. This assumption is closely approximated in
practice when the manipulated BDDs only occupy a small fraction of the
available nodes in the BDD package. Otherwise, performance may be improved by
increasing the BDD memory (see Subsection~\ref{ss:memory}).

When the operands of an expression are relations, the computation time is given
as function of the sizes of their BDD representation. (The only exception are
the transitive closure operators, where a function of the size of the universe
gives a more useful bound.) This raises the problem of how to estimate these
BDD sizes. Many practical relations have regularities that enable an (often
dramatically) compressed BDD representation, but the analytical derivation of
the typical compression rate for relations from a particular application domain
is generally difficult.  Our advice is to choose some representative examples
and measure the BDD sizes with the {\tt PRINT RELINFO} statement.

It is important to note that Table~\ref{t:complexity} gives {\em worst-case}
computation times.  In many cases, the typical practical performance is much
better than the worst case.  For example, the relational comparison operators
({\tt <=}, {\tt <}, {\tt >=}, {\tt >}) and the binary logic operators ({\tt
\&}, {\tt |}, {\tt ->}, {\tt <->}) are very common in RML programs. Their
worst-case complexity is the product of the sizes of their operand BDDs, which
is alarmingly high. However, in practice the performance is often much closer
to the sum of the operand BDD sizes. Similarly, the quantification operators
are often efficient despite their prohibitive worst case runtime (which is
difficult to derive because quantification is implemented as a series of
several bit-level operations).

Another practically important example for the gap between average-case and
worst-case runtime are the transitive closure operators.  The worst case
complexity of their BDD-based implementation is the same as for implementations
with conventional data structures. However, the BDD-based implementations are
much more efficient for many practical graphs~\cite{BeyerEtAl:2003}.  Even in
the comparison of different BDD-based implementations, a better worst-case
complexity does not imply a better performance in practice.  We conclude from
our experience that  knowledge of the theoretical complexity complements but
cannot replace experimentation in the development of highly optimized RML
programs.

\begin{table}
\vspace{-1ex}\centering

\normalsize
\begin{tabular}{l@{~~~~}l}
{\tt ! re}
    & $O(${\em bddsize}$(${\tt re}$))$ \\
{\tt re1 \& re2}, {\tt ~re1 | re2}
    & $O(${\em bddsize}$(${\tt re1}$)$ $\cdot$ {\em bddsize}$(${\tt re2}$))$\\
{\tt re1 -> re2}, {\tt ~re1 <-> re2}
    & $O(${\em bddsize}$(${\tt re1}$)$ $\cdot$ {\em bddsize}$(${\tt re2}$))$\\
{\tt EX(x, re)}, {\tt ~FA(x, re)}
    & $\ge O(${\em bddsize}$(${\tt re}$)^2)$ \\
{\tt TC(re)} & $O(n^3)$ \\
{\tt TCFAST(re)} & $O(n^3 \log n)$ \\
{\tt FALSE(x1, x2, ...)} & $O(1)$ \\
{\tt TRUE(x1, x2, ...)} & $O(1)$ \\
{\tt @s(x)} & $O(n \log n)$ \\
{\tt \symbol{126}(x1, x2)}
    & $O(n \log n)$ ({\tt \symbol{126}} can be
      {\tt =}, {\tt !=}, {\tt <=}, {\tt <}, {\tt >=}, {\tt >})\\
{\tt \symbol{126}(ne1, ne2)}
    & $O(1)$ ({\tt \symbol{126}} can be
      {\tt =}, {\tt !=}, {\tt <=}, {\tt <}, {\tt >=}, {\tt >})\\
{\tt re1 = re2}, {\tt ~re1 != re2} & $O(1)$ \\
{\tt re1 < re2}, {\tt ~re1 <= re2}
    & $O(${\em bddsize}$(${\tt re1}$)$ $\cdot$ {\em bddsize}$(${\tt re2}$))$\\
{\tt re1 > re2}, {\tt ~re1 >= re2}
    & $O(${\em bddsize}$(${\tt re1}$)$ $\cdot$ {\em bddsize}$(${\tt re2}$))$\\[2ex]
\end{tabular}

\caption{Worst case time complexity of the evaluation of relational
expressions. {\tt re}, {\tt re1}, {\tt re2} are relational expressions, {\tt
x}, {\tt x1}, {\tt x2} are attributes, and {\tt ne1}, {\tt ne2} are numerical
expressions. {\em bddsize}$(${\tt re}$)$ is the number of BDD nodes of the
result of the expression {\tt re}, and $n$~is the cardinality of the universe.}
\label{t:complexity}
\end{table}

\subsection{Extending the Universe} \label{s:universe}

The set of all strings that may be tuple elements of relations in an RML
program is called the {\em universe}.  The universe contains all tuple elements
of the input relations (from the input RSF data), and all string literals that
appear on the left hand side of assignments in the RML program.  The universe
is immutable in the sense that it can be determined before the interpretation
of the RML program starts, and is not changed during the interpretation of the
RML program.

Sometimes the immutability of the universe is inconvenient for the developer of
RML programs. Consider, for example, a program that takes as input the nodes
and arcs of a graph, and computes the binary relation {\tt OutneighborCnt}
which contains for each node the number of outneighbors:
\begin{verbatim}
FOR n IN Node(x) {
    OutneighborCnt(n, #(Arc(n,x))) := TRUE();
}
\end{verbatim}
This is not a syntactically correct RML program, because {\tt \#(Arc(n,x))} is
not a string, but a number. However, RML has an operator {\tt STRING} that
converts a number into a string.  But
\begin{verbatim}
OutneighborCnt(n, STRING( #(Arc(n,x)) )) := TRUE();
\end{verbatim}
is still not syntactically correct, because such a conversion is not allowed at
the left hand side of assignment statements. The reason is that the string that
results from such a conversion is generally not known before the execution of
the RML program, can therefore not be added to the universe before the
execution, and is thus not allowed as tuple element of a relation.

The immutability of the universe during the execution of an RML program is
necessary because constant relations like {\tt TRUE(x)} (the universe) and {\tt
=(x,y)} (string equality for all strings in the universe) are only defined for
a given universe.  Also, the complement of a relation depends on the universe:
The complement of a set contains all strings of the universe that are not in
the given set, and thus clearly changes when the universe changes.

However, there is a way to work around this limitation: Writing an RSF file,
and restarting CrocoPat with this RSF file as input, which adds all tuple
elements in the RSF file to the universe.  For example, the above incorrect
program can be replaced by the following correct program:
\begin{verbatim}
FOR n IN Node(x) {
    PRINT "OutneighborCnt ", n, " ", #(Arc(n,x)) TO "OutneighborCnt.rsf";
}
\end{verbatim}
When CrocoPat is restarted with the resulting RSF file {\tt OutneighborCnt.rsf}
as input, the binary relation {\tt OutneighborCnt} is available for further
processing.

\section{CrocoPat Reference} \label{s:pat}

CrocoPat is executed with
\begin{verbatim}
crocopat [OPTION]... FILE [ARGUMENT]...
\end{verbatim}
It first reads relations in RSF (see Section~\ref{s:rsf}) from stdin (unless
the option {\tt -e} is given) and then executes the RML program {\tt FILE} (see
Section~\ref{s:lan}).  The {\tt ARGUMENT}s are passed
to the RML program.  The {\tt OPTION}s are \\[1ex]
\begin{tabular}{l@{\hspace{3mm}}l}
{\tt -e}        &  Do not read RSF data from stdin. \\
{\tt -m NUMBER} &  Approximate memory for BDD package in MB. The default is~50. See Subsection~\ref{ss:memory}.\\
{\tt -q}        &  Suppress warnings. \\
{\tt -h}        &  Display help message and exit. \\
{\tt -v}        &  Print version information and exit. \\
\end{tabular}\\[1ex]
\pagebreak

The output of the RML program can be written to files, stdout, or stderr, as
specified in the RML program.  Error messages and warnings of CrocoPat are always written
to stderr.

The exit status of CrocoPat is~1 if it terminates abnormally and~0 otherwise.
CrocoPat always outputs an error message to stderr before it terminates with exit
status~1.

\section{RSF Reference} \label{s:rsf}

Rigi Standard Format (RSF) is CrocoPat's input and output format for relations.
It is an extension of the format for binary relations defined
in~\cite[Section~4.7.1]{Wong:1998}.  For examples of its use see
Subsection~\ref{s:output}.

An RSF stream is a sequence of lines.  The order of the lines is arbitrary. The
repeated occurrence of a line is permissable and has the same meaning as a
single occurrence. The end of an RSF stream is indicated by the end of the file
or by a line that starts with a dot ({\tt .}). Lines starting with a sharp
({\tt \#}) are comment lines.

All lines that do not start with a dot or a sharp specify a tuple in a
relation. They consist of the name of the relation followed by a sequence of
(an arbitrary number of) tuple elements, separated by at least one whitespace
character (i.e., space or horizontal tab).

Relation names must be RML identifiers (see Subsection~\ref{ss:lexical}). Tuple
elements are sequences of arbitrary characters except line breaks and
whitespace characters.  A tuple element may be optionally enclosed by double
quotes ({\tt "}), in which case it may also contain whitespace characters.
Tuple elements that are enclosed by double quotes in the RSF input of an RML
program are also enclosed by double quotes in its output.

\section{RML Reference} \label{s:lan}

Relation Manipulation Language (RML) is CrocoPat's programming language for
manipulating relations. This section defines the lexical structure, the syntax,
and the semantics of RML. Nonterminals are printed in italics and terminals in
typewriter.

\subsection{Lexical Structure} \label{ss:lexical}

Identifiers are sequences of Latin letters ({\tt a-zA-z}), digits ({\tt 0-9})
and underscores ({\tt \_}), the first of which must be a letter or underscore.
RML has four types of identifiers: attributes ({\em attribute}), relational
variables ({\em rel\_var}), string variables ({\em str\_var}), and numerical
variables ({\em num\_var}).  Every identifier of an RML program belongs to
exactly one of these types.  The type is determined at the first occurrence of
the identifier in the input RSF file (only possible for relational variables)
or in the RML program. Explicit declaration of identifiers is not necessary.

The following strings are reserved as keywords and therefore cannot be
used as identifiers:
\begin{verbatim}
AVG DIV ELSE ENDL EX EXEC EXIT FA FOR IF IN MAX MIN MOD NUMBER PRINT RELINFO
STDERR STRING SUM TC TCFAST TO WHILE
\end{verbatim}

RML has two types of literals: string literals ({\em str\_literal}) and
numerical literals ({\em num\_literal}).  String literals are delimited by
double quotes ({\tt "}) and can contain arbitrary characters except double
quotes.  A numerical literal consists of an integer part, a fractional part
indicated by a decimal point ({\tt .}), and an exponent indicated by the letter
{\tt e} or {\tt E} followed by an optionally signed integer.  All three parts
are optional, but at least one digit in the integer part or the fractional part
is required. Examples of numerical literals are {\tt 1}, {\tt .2}, {\tt 3.},
{\tt 4.5}, and {\tt 6e-7}.

There are two kinds of comments: Text starting with {\tt /*} and ending with
{\tt */}, and text from {\tt //} to the end of the line. \pagebreak

\subsection{Syntax and Informal Semantics}\label{s:syntax}

\footnotetext[1]{{\em stmt} ... denotes a sequence of one or more {\em stmt}s.}
\footnotetext[2]{Context conditions are marked with~$\to$.}
\footnotetext[3]{{\tt \symbol{126}} can be
       {\tt =}, {\tt !=}, {\tt <=}, {\tt <}, {\tt >=}, {\tt >}.}

\begin{tabbing}
\hspace*{52mm}\=\hspace{5mm}\= \kill
\noindent{\bfseries \em program} ::=
    \>\> {\bf RML program.}\\
{\em stmt} ...\footnotemark[1]
    \>\> Executes the {\em stmt}s in the given order.\\[1.5ex]

\noindent{\bfseries \em stmt} ::=
    \>\> {\bf Statement.}\\
{\em rel\_var}{\tt (}{\em term}{\tt ,}...{\tt )} {\tt := }{\em rel\_expr}{\tt ;}
    \>\> Assigns the result of {\em rel\_expr} to {\em rel\_var}.\\
    \>$\to$\footnotemark[2]\> The {\em term}s must be {\em attribute}s or {\em str\_literal}s. \\
    \>$\to$\> The set of {\em attribute}s among the {\em term}s on the left hand side \\
    \>\> must equal the set of free attributes in {\em rel\_expr}. \\
{\em rel\_var}{\tt (}{\em term}{\tt ,}...{\tt );}
    \>\> Shortcut for {\em rel\_var}{\tt (}{\em term}{\tt ,}...{\tt )} {\tt := TRUE(}{\em term}{\tt ,}...{\tt )}. \\
{\em str\_var}{\tt ~:= }{\em str\_expr}{\tt ;}
    \>\> Assigns the result of {\em str\_expr} to {\em str\_var}.\\
{\em num\_var}{\tt ~:= }{\em num\_expr}{\tt ;}
    \>\> Assigns the result of {\em num\_expr} to {\em num\_var}.\\
{\tt IF}~{\em rel\_expr}~{\tt \{}{\em stmt} ...{\tt\}}~{\tt ELSE}~{\tt \{}{\em stmt} ...{\tt\}}
    \>\> Executes the {\em stmt}s before {\tt ELSE} if the result of {\em rel\_expr} is {\tt TRUE()}, \\
{\tt IF}~{\em rel\_expr}~{\tt \{}{\em stmt} ...{\tt\}}
    \>\> and the {\em stmt}s after {\tt ELSE} (if present) otherwise.\\
    \>$\to$\> {\em rel\_expr} must not have free attributes.\\
{\tt WHILE }{\em rel\_expr}{\tt~\{}{\em stmt} ...{\tt\}}
    \>\> Exec. the {\em stmt}s repeatedly as long as {\em rel\_expr} evaluates to {\tt TRUE()}. \\
    \>$\to$\> {\em rel\_expr} must not have free attributes.\\
{\tt FOR~}{\em str\_var}{\tt~IN~}{\em rel\_expr}{\tt~\{}{\em stmt} ...{\tt\}}
    \>\> Executes the {\em stmt}s once for each element in the result of {\em rel\_expr}. \\
    \>$\to$\> {\em rel\_expr} must have exactly one free attribute.\\
{\tt PRINT }{\em print\_expr}{\tt ,}...{\tt ;}
    \>\> Writes the results of the {\em print\_expr}s to stdout. \\
{\tt PRINT }{\em print\_expr}{\tt ,}...{\tt~TO STDERR;}
    \>\> Writes the results of the {\em print\_expr}s to stderr. \\
{\tt PRINT }{\em print\_expr}{\tt ,}...{\tt~TO }{\em str\_expr}{\tt ;}
    \>\> Appends the results of the {\em print\_expr}s to the specified file.\\
{\tt EXEC }{\em str\_expr}{\tt ;}
    \>\> Executes the shell command given by {\em str\_expr}. \\
    \>\> The exit status is available as numerical constant {\tt exitStatus}.\\
{\tt EXIT }{\em num\_expr}{\tt ;}
    \>\> Exits CrocoPat with the given exit status. \\
{\tt \{ }{\em stmt} ...{\tt~\}}
    \>\> Executes the {\em stmt}s in the given order. \\[1.5ex]

\noindent{\bfseries \em rel\_expr} ::=
    \>\> {\bf Relational Expression. The result is a relation.} \\*
{\em rel\_var}{\tt (}{\em term}{\tt ,}...{\tt)}
    \>\> Atomic relational expression. \\*
{\em term~~rel\_var~~term}
    \>\> Same as {\em rel\_var}{\tt (}{\em term}{\tt ,} {\em term}{\tt)}. \\*
{\tt ! }{\em rel\_expr}
    \>\> Negation (not). \\*
{\em rel\_expr}{\tt~\& }{\em rel\_expr}
    \>\> Conjunction (and). \\*
{\em rel\_expr}{\tt~| }{\em rel\_expr}
    \>\> Disjunction (or). \\*
{\em rel\_expr}{\tt~-> }{\em rel\_expr}
    \>\> Implication (if). {\tt r1 -> r2 }is equivalent to{\tt~!(r1) | (r2)}.\\*
{\em rel\_expr}{\tt~<-> }{\em rel\_expr}
    \>\> Equivalence (if and only if). \\*
    \>\> {\tt r1 <-> r2 }is equivalent to{\tt~(r1 -> r2) \& (r2 -> r1)}.\\*
{\tt EX(}{\em attribute}{\tt ,}...{\tt , }{\em rel\_expr}{\tt )}
    \>\> Existential quantification of the {\em attribute}s. \\*
{\tt FA(}{\em attribute}{\tt ,}...{\tt , }{\em rel\_expr}{\tt )}
    \>\> Universal quantification of the {\em attribute}s. \\*
{\tt TC(}{\em rel\_expr}{\tt )}
    \>\> Transitive closure. \\*
    \>$\to$\> {\em rel\_expr} must have exactly two free attributes. \\*
{\tt TCFAST(}{\em rel\_expr}{\tt )}
    \>\> Same as {\tt TC}, but with an alternative algorithm (see Section~\ref{s:control}). \\*
{\tt FALSE(}{\em term}{\tt ,}...{\tt )}
    \>\> Empty relation. \\*
{\tt TRUE(}{\em term}{\tt ,}...{\tt )}
    \>\> Relation containing all tuples of strings in the universe. \\*
{\tt @}{\em str\_expr}{\tt (}{\em term}{\tt )}
    \>\> Strings in the universe that match the regular expression {\em str\_expr}. \\*
{\tt \symbol{126}(}{\em term}{\tt , }{\em term}{\tt)}
    \>\> Lexicographical order of all strings in the universe.\footnotemark[3] \\*
{\em term}{\tt~\symbol{126} }{\em term}
    \>\> Same as {\tt \symbol{126}(}{\em term}{\tt , }{\em term}{\tt)}.\\*
{\tt \symbol{126}(}{\em num\_expr}{\tt , }{\em num\_expr}{\tt)}
    \>\> Numerical comparison. The result is either {\tt TRUE()} or {\tt FALSE()}.\footnotemark[3] \\*
{\em num\_expr}{\tt~\symbol{126} }{\em num\_expr}
    \>\> Same as {\tt \symbol{126}(}{\em num\_expr}{\tt , }{\em num\_expr}{\tt)}.\\*
{\em rel\_expr}{\tt~\symbol{126} }{\em rel\_expr}
    \>\> Relational comparison. The result is either {\tt TRUE()} or {\tt FALSE()}.\footnotemark[3] \\*
{\tt (}{\em rel\_expr}{\tt )}
\end{tabbing}

\begin{tabbing}
\hspace*{45mm}\=\hspace{5mm}\= \kill
\noindent{\bfseries \em term} ::=
    \>\> {\bf Term.} \\
{\em attribute}
    \>\> Attribute. \\
{\tt \_}
    \>\> Anonymous attribute. E.g. {\tt R(\_)} is equivalent to {\tt EX(x, R(x))}.\\
{\em str\_expr}
    \>\> String expression. \\[2ex]

\noindent{\bfseries \em str\_expr} ::=
    \>\> {\bf String Expression. The result is a string.} \\
{\em str\_literal}
    \>\> String literal.\\
{\em str\_var}
    \>\> String variable.\\
{\tt STRING(}{\em num\_expr}{\tt )}
    \>\> Converts the result of {\em num\_expr} into a string. \\
{\tt \$ }{\em num\_expr}
    \>\> Command line argument. The first argument has the number~$1$. \\
    \>\> The constant {\tt argCount} contains the number of arguments. \\
{\em str\_expr}{\tt~+ }{\em str\_expr}
    \>\> Concatenation. \\
{\tt (}{\em str\_expr}{\tt )}\\[2ex]

\noindent{\bfseries \em num\_expr} ::=
    \>\> {\bf Numerical Expression. The result is a floating point number.} \\
{\em num\_literal}
    \>\> Numerical literal.\\
{\em num\_var}
    \>\> Numerical variable.\\
{\tt NUMBER(}{\em str\_expr}{\tt )}
    \>\> Converts the result of {\em str\_expr} into a number. Yields~$0.0$ if \\
    \>\> the result of {\em str\_expr} is not the string representation of a number.\\
{\tt \#(}{\em rel\_expr}{\tt )}
    \>\> Cardinality (number of elements) of the result of {\em rel\_expr}. \\
{\tt MIN(}{\em rel\_expr}{\tt )}, {\tt MAX(}{\em rel\_expr}{\tt )}, {\tt SUM(}{\em rel\_expr}{\tt )}, {\tt AVG(}{\em rel\_expr}{\tt )} \\
    \>\> Minimum, maximum, sum, and arithmetic mean of {\tt NUMBER(s)} \\
    \>\> over all strings {\tt s} in the result of {\em rel\_expr}. \\
    \>$\to$\> {\em rel\_expr} must have one free attribute, its result must be non-empty. \\
{\em num\_expr}{\tt~+ }{\em num\_expr}
    \>\> Addition. \\
{\em num\_expr}{\tt~- }{\em num\_expr}
    \>\> Subtraction. \\
{\em num\_expr}{\tt~* }{\em num\_expr}
    \>\> Multiplication. \\
{\em num\_expr}{\tt~/ }{\em num\_expr}
    \>\> Real division. \\
{\em num\_expr}{\tt~DIV }{\em num\_expr}
    \>\> Integer division (truncating). \\
{\em num\_expr}{\tt~MOD }{\em num\_expr}
    \>\> Modulo. \\
{\em num\_expr}{\tt~\symbol{94} }{\em num\_expr}
    \>\> The first {\em num\_expr} raised to the power given by the second {\em num\_expr}. \\
{\tt argCount}
    \>\> Number of command line arguments. \\
{\tt exitStatus}
    \>\> Exit status of the last executed {\tt EXEC} statement. \\
{\tt (}{\em num\_expr}{\tt )}\\[2ex]

\noindent{\bfseries \em print\_expr} ::=
    \>\> {\bf Print Expression.} \\
{\em rel\_expr}
    \>\> Prints the tuples in the result of {\em rel\_expr}, one tuple per line.\\
{\tt [}{\em str\_expr}{\tt ] }{\em rel\_expr}
    \>\> Prints the result of {\em str\_expr} before each tuple of {\em rel\_expr}. \\
{\em str\_expr}
    \>\> Prints the result of {\em str\_expr}.\\
{\em num\_expr}
    \>\> Prints the result of {\em num\_expr}.\\
{\tt ENDL}
    \>\> Prints a line break. \\
{\tt RELINFO(}{\em rel\_expr}{\tt )}
    \>\> Prints information about the BDD representation of {\em rel\_expr}.
\end{tabbing}

\noindent{\bf Levels of Precedence (from Low to High)}\\
{\tt =}, {\tt !=}, {\tt <=}, {\tt <}, {\tt >=}, {\tt >} \\
{\tt ->}, {\tt <->}\\
{\tt |}\\
{\tt \&}\\
{\tt !}\\
{\tt +}, {\tt -} (binary) \\
{\tt *}, {\tt /}, {\tt DIV}, {\tt MOD} \\
{\tt \symbol{94}}\\
{\tt -} (unary)\\
{\tt \$}

\subsubsection{Free Attributes in Relational Expressions.}
Several context conditions of RML refer to the free attributes in relational
expressions.  The number of free attributes in a relational expression equals
the arity of the resulting relation.  Informally, the set of free attributes of
an expression is the set of its contained attributes that are not in the scope
of a quantifier (i.e., {\tt EX} or {\tt FA}).  Exceptions are the numerical and
relational comparison, which have Boolean results and therefore no free
attributes. Formally, the function~$f$ that assigns to each relational
expression the set of its free attributes is inductively defined as
follows:\\[1ex]
\begin{tabular}{l@{\hspace{2mm}}l@{\hspace{2mm}}l}
$f($~{\em rel\_var}{\tt (}{\em term}$_1${\tt ,} {\em term}$_2${\tt ,}...{\tt)}$~)$
    & $:=$ & $\{$ {\em term}$_i~|~${\em term}$_i$ is an attribute $\}$ \\
$f($~{\tt !}{\em rel\_expr}~$)$
    & $:=$ & $f($~{\em rel\_expr}~$)$\\
$f($~{\em rel\_expr}$_1${\tt~\& }{\em rel\_expr}$_2$~$)$
    & $:=$ & $f($ {\em rel\_expr}$_1$~$) ~\cup~ f($~{\em rel\_expr}$_2$~$)$ \\
$f($~{\em rel\_expr}$_1${\tt~| }{\em rel\_expr}$_2$~$)$
    & $:=$ & $f($ {\em rel\_expr}$_1$~$) ~\cup~ f($~{\em rel\_expr}$_2$~$)$ \\
$f($~{\em rel\_expr}$_1${\tt~-> }{\em rel\_expr}$_2$~$)$
    & $:=$ & $f($ {\em rel\_expr}$_1$~$) ~\cup~ f($~{\em rel\_expr}$_2$~$)$ \\
$f($~{\em rel\_expr}$_1${\tt~<-> }{\em rel\_expr}$_2$~$)$
    & $:=$ & $f($ {\em rel\_expr}$_1$~$) ~\cup~ f($~{\em rel\_expr}$_2$~$)$ \\
$f($~{\tt EX(}{\em attribute}{\tt , }{\em rel\_expr}{\tt )}~$)$
    & $:=$ & $f($ {\em rel\_expr} $) ~\setminus~ \{$ {\em attribute} $\}$\\
$f($~{\tt FA(}{\em attribute}{\tt , }{\em rel\_expr}{\tt )}~$)$
    & $:=$ & $f($ {\em rel\_expr} $) ~\setminus~ \{$ {\em attribute} $\}$\\
$f($~{\tt TC(}{\em rel\_expr}{\tt )}~$)$
    & $:=$ & $f($~{\em rel\_expr}~$)$\\
$f($~{\tt TCFAST(}{\em rel\_expr}{\tt )}~$)$
    & $:=$ & $f($~{\em rel\_expr}~$)$\\
$f($~{\tt (}{\em num\_expr$_1$}{\tt~\symbol{126} }{\em num\_expr$_2$}{\tt )}~$)$
    & $:=$ & $\emptyset$ \\
$f($~{\tt (}{\em rel\_expr$_1$}{\tt~\symbol{126} }{\em rel\_expr$_2$}{\tt )}~$)$
    & $:=$ & $\emptyset$ \\
$f($~{\tt (}{\em rel\_expr}{\tt )} $)$
    & $:=$ & $f($~{\em rel\_expr}~$)$
\end{tabular} \\[1ex]
As before, {\tt\symbol{126}} can be {\tt =}, {\tt !=}, {\tt <=}, {\tt <}, {\tt
>=}, or {\tt >}. The relational constants {\tt\symbol{126}}, {\tt FALSE},
{\tt TRUE}, and {\tt @}{\em str\_expr} are equivalent to {\em rel\_var} with
respect to the definition of free attributes.

\subsubsection{Regular Expressions.} In the relational expression
{\tt @}{\em str\_expr}{\tt (}{\em term}{\tt )}, the result of {\em str\_expr}
can be any POSIX extended regular expression~\cite[Section~9.4]{IEEE:2001}.  A
full description is beyond the scope, we only give a short overview.

Most characters in a regular expression only match themselves.  The following
special characters match themselves only when they are preceded by a
backslash~({\tt \symbol{92}}), and otherwise have special meanings:\\[1ex]
\begin{tabular}{ll}
{\tt .}  & Matches any single character.\\
$\mathtt{[~]}$ & Matches any single character contained within the brackets.\\
$\mathtt{[}${\tt \symbol{94}}~$\mathtt{]}$ & Matches any single character not contained within the
brackets.\\
{\tt \symbol{94}} & Matches the start of the string.\\
{\tt \$} & Matches the end of the string.\\
{\tt \{$x$,$y$\}}~~ & Matches the last character (or regular expression enclosed by
parentheses) \\
                    & at least $x$ and at most $y$ times.\\
{\tt +}  & Matches the last character (or regular expression enclosed by
parentheses) one or more times.\\
{\tt *}  & Matches the last character (or regular expression enclosed by
parentheses) zero or more times.\\
{\tt ?}  & Matches the last character (or regular expression enclosed by
parentheses) zero or one times.\\
{\tt |}  & Matches either the expression before or the expression after the
operator.
\end{tabular}\\[1ex]
Regular expressions can be grouped by enclosing them with
parentheses~({\tt(}...{\tt)}).

\subsection{Formal Semantics}\label{s:semantics}

This subsection formally defines the semantics of the part of RML that deals
with relations, namely of relational expressions and the relational assignment
statement.

The {\em universe} $\cc$ is the finite set of all string literals that appear
in the input RSF file, or on the left side of a relational assignment. The
finite set of {\em attributes} of the RML program is denoted by $\cx$ ($\cc
\cap \cx = \emptyset$). An {\em attribute assignment} is a total function $v:
\cx \cup \cc \to \cc$ which maps each attribute to its value and (for
notational convenience) each string literal to itself.  The set of all
attribute assignments is denoted by $\mathit{Val(\cx)}$.

The finite set of {\em relation variables} in the RML program is denoted by
$\crel$.  A {\em relation assignment} is a total function $s: \crel \to
\bigcup_{n \in \mathbb{N}} 2^{\prod_{k=1}^n \cc}$, which maps each relation
variable to a relation of arbitrary arity. The set of all relation assignments
is denoted by~$Rel(\crel)$.

The semantics of relational expressions and statements are given
by the following interpretation functions:
\[
\begin{array}{lcll}
\bbbkl.\bbbkr_e &:& \mbox{\em rel\_expr}s \to
                  ( Rel(\crel) \to 2^{\mathit{Val}(\cx)} ) \\[1ex]
\bbbkl.\bbbkr_s &:& \mbox{\em stmt}s \to
                  ( Rel(\crel) \to Rel(\crel) ) \\
\end{array}
\]
So we define the semantics of an expression as the set of attribute assignments
that satisfy the expression, and the semantics of a statement as a
transformation of the relation assignment. The interpretation functions are
defined inductively in Figure~\ref{fig:sem}.

\begin{figure*}
\[
\hspace{-5mm}  \begin{array}{rcll}
    \bbbkl \mbox{{\em rel\_var}{\tt (}{\em term}$_1${\tt ,}~$\dots${\tt ,}~{\em term}$_n${\tt )}} \bbbkr_e (s) &=&
      \big\{ v \in Val(\cx) ~\big|~
        \big( v(term_1), \dots, v(term_n) \big) \in s(\mbox{\em rel\_var})
      \big\}\\[1ex]

    \bbbkl \mbox{{\tt !}~{\em rel\_expr}} \bbbkr_e (s) &=&
      \mathit{Val}(\cx) \setminus \bbbkl \mbox{{\em rel\_expr}} \bbbkr_e (s) \\[1ex]

    \bbbkl \mbox{{\em rel\_expr$_1$}~{\tt \&}~{\em rel\_expr$_2$}} \bbbkr_e (s) &=&
      \bbbkl \mbox{{\em rel\_expr$_1$}} \bbbkr_e(s) \cap
      \bbbkl \mbox{{\em rel\_expr$_2$}} \bbbkr_e(s) \\[1ex]

    \bbbkl \mbox{{\em rel\_expr$_1$}~{\tt |}~{\em rel\_expr$_2$}} \bbbkr_e (s) &=&
      \bbbkl \mbox{{\em rel\_expr$_1$}} \bbbkr_e(s) \cup
      \bbbkl \mbox{{\em rel\_expr$_2$}} \bbbkr_e(s) \\[1ex]

    \bbbkl \mbox{{\em rel\_expr$_1$}~{\tt ->}~{\em rel\_expr$_2$}} \bbbkr_e (s) &=&
      \bbbkl \mbox{{\tt !}~{\tt (}{\em rel\_expr$_1$}{\tt )}~{\tt |\,(}{\em rel\_expr$_2$}{\tt )}} \bbbkr_e (s) \\[1ex]

    \bbbkl \mbox{{\em rel\_expr$_1$}~{\tt <->}~{\em rel\_expr$_2$}} \bbbkr_e (s) &=&
      \bbbkl \mbox{{\tt (}{\em rel\_expr$_1$}~{\tt ->}~{\em rel\_expr$_2$}{\tt )}~{\tt \&\,}{\tt (}{\em rel\_expr$_2$}~{\tt ->}~{\em rel\_expr$_1$}{\tt )}} \bbbkr_e (s) \\[1ex]

    \bbbkl \mbox{{\tt EX(}{\em attribute}{\tt ,}~{\em rel\_expr}{\tt )}} \bbbkr_e (s) &=&
      \big\{ v \in Val(\cx) ~\big|~ \exists v' \in \bbbkl \mbox{{\em rel\_expr}} \bbbkr_e(s)
                                   ~\forall x \in \cx \setminus \{\mbox{{\em attribute}}\}: v(x) = v'(x) \big\}\\[1ex]

    \bbbkl \mbox{{\tt FA(}{\em attribute}{\tt ,}~{\em rel\_expr}{\tt )}} \bbbkr_e (s) &=&
      \bbbkl \mbox{{\tt !}~{\tt EX(}{\em attribute}{\tt ,}~{\tt !}~{\tt (}{\em rel\_expr}{\tt ))}} \bbbkr_e (s) \\[1ex]

    \bbbkl \mbox{{\tt TC(}{\em rel\_var}{\tt (}{\em attribute}$_1${\tt ,}~{\em attribute}$_2${\tt ))}} \bbbkr_e (s)
                &\stackrel{\mbox{\em \footnotesize lfp}}{=}&
      \bbbkl \mbox{~~~~{\em rel\_var}{\tt (}{\em attribute}$_1${\tt ,}~{\em attribute}$_2${\tt )}} \\
         &&\mbox{~~{\tt | EX(}{\em attribute}$_3${\tt ,}~~~
           {\em rel\_var}{\tt (}{\em attribute}$_1${\tt ,}~{\em attribute}$_3${\tt )}}\\
         &&\mbox{\hspace{27mm}{\tt \& TC(}{\em rel\_var}{\tt (}{\em attribute}$_3${\tt ,}~{\em attribute}$_2${\tt )))}} \bbbkr_e (s) \\[1ex]

    \bbbkl \mbox{{\tt TCFAST(}{\em rel\_var}{\tt (}{\em attribute}$_1${\tt ,}~{\em attribute}$_2${\tt ))}} \bbbkr_e (s) &=&
      \bbbkl \mbox{{\tt TC(}{\em rel\_var}{\tt (}{\em attribute}$_1${\tt ,}~{\em attribute}$_2${\tt ))}} \bbbkr_e (s) \\[3ex]

    \bbbkl \mbox{{\tt TRUE(}{\em term}$_1${\tt ,}~$\dots${\tt ,}~{\em term}$_n${\tt )}} \bbbkr_e (s) &=&
      \mathit{Val}(\cx)\\[1ex]

    \bbbkl \mbox{{\tt FALSE(}{\em term}$_1${\tt ,}~$\dots${\tt ,}~{\em term}$_n${\tt )}} \bbbkr_e (s) &=&
      \emptyset \\[1ex]

    \bbbkl \mbox{{\tt =(}{\em term}$_1${\tt ,}~{\em term}$_2${\tt )}} \bbbkr_e (s) &=&
      \big\{ v \in Val(\cx) ~|~ v(term_1) = v(term_2) \big\}\\[1ex]

    \bbbkl \mbox{{\tt <(}{\em term}$_1${\tt ,}~{\em term}$_2${\tt )}} \bbbkr_e (s) &=&
      \big\{ v \in Val(\cx) ~|~ v(term_1) <_{\mathrm{lexicographically}} v(term_2) \big\}\\[3ex]

    \bbbkl \mbox{{\tt =(}{\em rel\_expr}$_1${\tt ,}~{\em rel\_expr}$_2${\tt )}} \bbbkr_e (s) &=&
      \mathit{Val}(\cx) \mbox{,~~~if }
        \bbbkl \mbox{{\em rel\_expr$_1$}} \bbbkr_e(s) = \bbbkl \mbox{{\em rel\_expr$_2$}} \bbbkr_e(s) \\
      && \emptyset \mbox{,~~~~~~~~~~~otherwise}\\[1ex]

    \bbbkl \mbox{{\tt <(}{\em rel\_expr}$_1${\tt ,}~{\em rel\_expr}$_2${\tt )}} \bbbkr_e (s) &=&
      \mathit{Val}(\cx) \mbox{,~~~if }
        \bbbkl \mbox{{\em rel\_expr$_1$}} \bbbkr_e(s) \subsetneq \bbbkl \mbox{{\em rel\_expr$_2$}} \bbbkr_e(s) \\
      && \emptyset \mbox{,~~~~~~~~~~~otherwise}\\[1ex]

    \bbbkl \mbox{{\tt (}{\em rel\_expr}{\tt )}} \bbbkr_e(s) &=&
      \bbbkl \mbox{\em rel\_expr} \bbbkr_e(s)\\[3ex]

    \big( \bbbkl \mbox{{\em rel\_var}{\tt (}{\em term}$_1${\tt ,}$\dots${\tt ,}{\em term}$_n${\tt )}{\tt :=}~{\em rel\_expr}} \bbbkr_s (s)\big) (r) &=&
      s(r), \hspace{64mm}\mbox{if } r \not= \mbox{\em rel\_var}\\[1.5ex]
      && \quad \big\{ \big( v(term_1), \dots, v(term_n) \big)~\big|~v \in \bbbkl \mbox{\em rel\_expr} \bbbkr_e(s) \big\} \\[1ex]
      && \cup~\big\{ t \in s(\mbox{\em r})~|~ \exists i : term_i \in \cc \wedge term_i \not= t_i \big\},
         \qquad\quad \mbox{if } r = \mbox{\em rel\_var}
  \end{array}
\]
with $\mbox{\em attribute, attribute$_1$, attribute$_2$, attribute$_3$} \in \cx
\mbox{, \em term$_1$, ..., term$_n$} \in \cx \cup \cc \mbox{, and \em rel\_var}
\in \crel$.  The symbol $\stackrel{\mbox{\em \footnotesize lfp}}{=}$ denotes
the least fixed point.

\caption{RML semantics} \label{fig:sem}
\end{figure*}

\newcommand{\etalchar}[1]{$^{#1}$}

\end{document}